\documentclass[]{revtex4-2}
\usepackage{ulem}
\usepackage{graphicx}
\usepackage{comment}
\usepackage{verbatim}
\usepackage{dcolumn}
\usepackage{bm}
\usepackage[utf8]{inputenc}
\usepackage[compat=1.1.0]{tikz-feynman}
\usepackage{bbm}
\usepackage{amsmath}

\begin{document}

\preprint{APS/123-QED}

\title{Rough neutron fields and nuclear reactor noise}

\author{Eric Dumonteil}
\affiliation{
 Institut de Recherche sur les Lois Fondamentales de l'Univers\\
 CEA, Université Paris-Saclay, 91191 Gif-sur-Yvette, France
}

\date{\today}

\begin{abstract}
 
Nuclear reactor cores achieve sustained fission chain reactions through the so-called 'critical state' -a subtle equilibrium between their material properties and their geometries. Observed at macroscopic scales during operations, the resulting stationary neutron field is tainted by a noise term, that hinders various fluctuations occurring at smaller scales. These fluctuations are either of a stochastic nature (whenever the core is operated at low power) or related to various perturbations and vibrations within the core, even operated in its power regime. For reasons that are only partially understood using linear noise theory, incidental events have been reported, characterized by an increase of the power noise. Such events of power noise growth, sometimes up to seemingly unbounded levels, have already led in the past to volontary scramming of reactors.

In this paper, we will extend the findings of \cite{percolation_pre} (where a statistical field theory of critical processes was employed to model stochastic neutron noise) by incorporating the effects of power noise. We will show that the evolution of the neutron field in a reactor is intimately connected to the dynamic of surface growths given by the Kardar-Parisi-Zhang equation. Recent numerical results emerging from renormalization group approaches will be used to calculate a threshold in the amplitude of the reactor noise above which the core could enter a new criticality state, and to estimate the critical exponents characterizing this phase transition to rough neutron fields.  The phenomenology of this roughening transition will be correlated and compared to data of misunderstood reactor noise levels and reactor instabilities, and will be shown to provide both qualitative and quantitative insights into this long-standing issue of reactor physics.

\end{abstract}

\maketitle

\section{\label{sec:intro} Introduction }

Neutron transport theory aims at studying the behavior of neutron gaz evolving in multiplicative media. It is key to reactor physics, as it constitutes the foundation upon which simulation codes are developed and used to support all industrial activities revolving around reactors, such as radiation shielding and criticality safety studies, fuel cycle management, and reactor operations, just to name a few. 
It supposes that the average behavior of the dilute neutron gas is faithfully described by a linear Boltzmann equation, clearing the path to the development and use of various approximation schemes. The most widespread of these schemes is the so-called diffusion theory: neutrons are considered at thermal equilibrium so that the spatiotemporal characteristics of their population -and the associated core power distribution- are properly reproduced by a simple heat equation \footnote{sometimes the two-groups diffusion theory is also used, that couples two diffusion equations in two energy domains as to grasp spectral phenomena}. With proper adjustments of the coefficients of diffusion equations, the numerical schemes are so accurate that they are implemented in the automatic protection systems of many commercial reactors for core operations and scramming procedures.

However, power fluctuations can affect the overall statistics of neutrons within the reactor. At start-up, for example, the so-called zero-power state of the core can be subjected to stochastic fluctuations, arising from the random path of neutron flights and from the stochastic nature of the fission process, producing variable number of neutrons each time heavy nuclei are fissioned. The fission process in fact equips the random walk with a branching structure \cite{harris_book, Zoia2011ResidenceRevisited, Zoia2011CollisionFlights}. Building on a number of theoretical studies \cite{Dumonteil2014ParticleSimulations,Zoia2014ClusteringGeometries,Sutton2017NeutronCalculations}, recent experimental results \cite{Dumonteil2021PatchyReactions} have provided a concrete example of such low-power phenomena developing in the core: the emergence of spatio-temporal correlations, dubbed 'neutron clustering', offers a concrete example of an in situ mechanism ultimately leading to deviations from the expected neutron distribution.

At high reactor power, the need to cool the core imposes high coolant flow rates, ultimately leading to various vibrations of its components such as the internals, the fuel rods, or the assemblies. The water serving as a neutron moderator, these vibrations therefore translate in a random noise on local reactivities (i.e. related to the net production of neutrons) because induced fissions are extremely sensitive to the incoming neutron energies. The observable consequence is that the neutron flux, measured by in-core or ex-core detectors, also develops fluctuations around its mean value. Being used as a diagnostic tools to characterize and eventually locate reactor core perturbations \cite{Demaziere2004DevelopmentRinghals-2}, neutron noise is also perceived as a safety concern. Indeed, on the one hand, reactor automatic protection systems calculate in real-time a derivative of the flux to appreciate its variations. This enables detection, for instance, of accidental power transients: blinding such a protection system may jeopardize the safety strategy. On the other hand, more worryingly, high reactor noise levels and instabilities sometimes appear: power fluctuations and oscillations progressively or abruptly might increase to require, ultimately, reactor scramming procedures \cite{Torres2019NeutronCore}. Such events have attracted a great deal of attention in the last 50 years to the present day \cite{Verma2020AssessmentReactor} and have triggered theoretical \cite{2010HandbookEngineering} and numerical \cite{Belanger2021ExactProblems} investigations. Although known to be associated with fluctuations in coolant temperature and vibrations of reactor components, these events and the parameters that influence them are still only partially understood \cite{Seidl2015ReviewPWRs}. 

Although formal approaches to describing fluctuations are generally provided in the context of linear noise theory, recent developments in the statistical field theory of reaction-diffusion processes \cite{garcia_millan} paved the way to describing the stochastic fluctuations of the branching random walk using critical phenomena: in particular, in the presence of stabilizing feedbacks (such as the Doppler effect \footnote{when the temperature in the fuel increases, the Doppler broadening of neutron cross-sections tends to increase capture resonances and hence constitutes one of the main feedback mechanism}), these fluctuations were proved to be related to a time-directed percolation process and to shift the effective random walk of neutrons from a diffusive to a super-diffusive behavior \cite{percolation_pre}. While such counter-reactions do not occur at zero power, this modeling described, however, very well the behavior of simulated population of neutrons: implementing population control mechanisms \cite{DeMulatier2015TheRevisited,Sutton2017NeutronCalculations}, Monte Carlo simulations of reactors (even when deployed on peta-scale architectures) imply neutron statistics only met during the first stages of reactor startup.

This paper now addresses the modeling of real neutron noise occurring in the power regime of reactors, using a statistical field theory of critical processes. In Section \ref{sec:diffusion} we will describe the statistics of neutrons in fissile media using a stochastic heat equation with multiplicative noise. Relating this equation to the celebrated Kardar-Parisi-Zhang (KPZ) equation, we will show that the reactor might be subjected to a roughening phase transition, where the neutron field becomes rough and is described by the exponents of the KPZ universality class, even when reactor feedbacks are considered. Using results from both perturbative and non-perturbative numerical approaches to estimate both the phase transition threshold and the fixed point value, Section \ref{sec:noise} will first briefly present the phenomenology of reactor instabilities and then select two reactor instabilities events so as to correlate the roughening transition model with reactor experimental data.


\section{\label{sec:diffusion} Nuclear reactor model and the KPZ equation}

At start-up, the power in nuclear reactors can be extremely small -around 1 kW-, and the associated neutron statistics is therefore close to $~10^{6}$ cm$^{-3}$, lending itself to stochastic fluctuations. The power regime is instead associated to high neutron densities: indeed, when the reactor reaches its nominal power, the neutron densities are $~10^{13}$ cm$^{-3}$ and stochastic fluctuations have been totally washed out. At this point, fission reactions release gigawatts of thermal power : such a heat source requires to be transferred from the primary to the secondary loop, through increased flow rates of the coolant primary pumps. The subsequent temperature fluctuations and induced vibrations are therefore intrinsically associated to high neutron densities and high fuel temperatures, and feedback effects might need to be taken into account for a proper description of the neutron gas. In this section, we will first build a simple reactor model in the intermediate temperature regime -where thermal feedback mechanisms can safely be neglected-, and then in the high temperature regime, adding a quadratic field to model the Doppler temperature effect.

\subsection{The nuclear reactor model}
\label{sub:nrm}

Industrial schemes used to operate light-water reactors often rely on the two-group diffusion approximation: thermal neutrons induce fissions and outgoing neutrons belong to a fast energy group prior to their slowing down to the thermal group. Both groups are simple diffusion equations with proper coefficients. Even a one-group diffusion equation allows to grasp the main spatial features of the neutron field. The temporal behavior is also adequately described by a double time scale: most neutrons produced during fissions are called prompt neutrons and are emitted on a short ($10^{-14}$ s) time scale, while a small fraction is produced following electroweak interactions within fission fragments on rather long time scales (from 1 s to 1 min). However, the study of stationary spatial distribution can safely be based on only one effective time scale, denoted $\Lambda$. Finally, as is usually done, the resulting one-group diffusion equation is supplemented by a stochastic term that models the random temporal and spatial fluctuations within the core. The resulting equatio can be written as
\begin{equation}
    \label{eq_diff} 
    \frac{\partial}{\partial t} n(\vec{x},t) =  D \Delta n(\vec{x},t) + \frac{\rho}{\Lambda} n(\vec{x},t) + \sigma \xi(\vec{x},t) n(\vec{x},t),
\end{equation}
where $n(\vec{x},t)$ denotes the neutron density field, $D$ is the diffusion coefficient, $\rho$ is the dimensionless reactivity (a net balance between neutron productions by fissions and neutron disappearance by captures), and where $\sigma \xi(\vec{x},t)$ is an uncorrelated white noise of strength $\sigma^2$ given by 
\begin{equation}
    \langle \xi(\vec{x},t) \xi(\vec{x}',t') \rangle  =\delta^{3}(\vec{x}-\vec{x}') \delta(t-t').
\end{equation}
The multiplicative white noise term reflects the various periodic fluctuations in the power regime (see, for instance, Ref.~\cite{williams_book} among many others), that occur on different spatial scales: for instance, in pressurized water reactors (PWR) the smallest perturbation size $a$ is such that $a \sim 1$ cm, and is associated to temperature fluctuations in the coolant flow \cite{Zylbersztejn2013OnPWRs}, while mechanical vibrations of the core components range from intermediate scales up to the whole reactor typical size $L$ -between two and three orders of magnitude above. In fact, both scales introduce natural regularization of the white noise: $a$ in the UV sector and $L$ in the IR sector. Noticeably, stochastic fluctuations occurring on microscopic scales cannot be adequately grasped by such a multiplicative noise model but can instead be described by additive noise \cite{percolation_pre}. 

\subsection{The roughening transition model}

\textbf{Reactor power noise and dynamic of surface growths}
Known in statistical physics literature as the Stochastic Heat Equation (SHE) with multiplicative noise, Eq.~\ref{eq_diff} can be reformulated using the following change of variables
\begin{align}
\label{paramSHEKPZ}
\begin{split}
    & \nu = D, \\
    & \eta = \frac{2 \nu}{\lambda} \sigma, \\
    & c = 2 D \frac{\rho/\Lambda}{\lambda}, \\
\end{split}
\end{align}
where a free positive parameter $\lambda$ was introduced. Then, associating the neutron density field $n(\vec{x},t)$ to a height field $h(\vec{x},t)$ through a Cole-Hopf transformation $n(\vec{x},t) = \exp[ \frac{\lambda}{2 \nu } h(\vec{x},t) ]$ allows to map our stochastic heat equation to a $\lambda$-parameterized family of Kardar-Parisi-Zhang (KPZ) equations supplemented by an additive term $c$
\begin{equation}
\label{kpz_eq}
    \frac{\partial}{\partial t} h(\vec{x},t) = \nu \Delta h + \frac{\lambda}{2} \left( \vec{\nabla} h \right)^2 + \eta \, \xi(\vec{x},t) + c.
\end{equation}
Whenever the reactor is stationary, $\rho$ is close to 0 and the constant term $c$ also becomes small. Conversely, the KPZ equation is mapped onto the SHE through
\begin{align}
\label{paramKPZSHE}
\begin{split}
    & \rho/\Lambda =  \frac{\lambda c}{2 \nu},\\
    & D = \nu, \\
    & \sigma = \frac{\lambda}{2 \nu} \eta. \\
\end{split}
\end{align} 
This connexion between the power noise equation of reactor physics and the KPZ equation, which relates the power fluctuations of a reactor to the dynamics of surface growths, went largely unnoticed in the reactor physics community. To better understand it, and since the choice of a particular Cole-Hopf transformation fixes the relative amount of stochasticity and of nonlinearity in the KPZ equation, it is possible to choose the natural gauge $\lambda/\nu=2$ which leads to the equation
\begin{equation}
\label{kpz_eq2}
    \frac{\partial}{\partial t} h(\vec{x},t) = D \Delta h + D \left( \vec{\nabla} h \right)^2 + \sigma \, \xi(\vec{x},t) + \rho/\Lambda.
\end{equation}
This particular form underlines a simple alphabet connecting the neutron density field to surface growth. In fact, the speed of surface growth is given by the reactivity state of the reactor: when the reactor is supercritical with constant positive reactivity $\rho$, the surface grows at a speed $\rho/ \Lambda$. Also, while the neutron diffusion coefficient $D$ drives both the smoothering by evaporation and the lateral growth of the surface in the same proportions (in this particular gauge), the multiplicative noise of the reactor in the power regime becomes an additive noise in the KPZ equation. In a spirit of generality, in the following we will not fix any particular gauge and use Eq.~\ref{paramSHEKPZ} and Eq.~\ref{paramKPZSHE} to connect the KPZ equation and the SHE equation.\\

\textbf{KPZ renormalization in dimension $d=3$}
Surprisingly, when looking at characterizing the KPZ universality class, it is customary to use the inverse Cole-Hopf transformation so as to get ride of $\lambda$ nonlinearity in the KPZ equation. In this context, the multiplicative noise equations describing reactor fluctuations are used, in a mirror approach, as a mean to solve surface growth equations. Merely looking for briefly summarizing a large literature on the subject, the study of fluctuations in statistical field theory can be decomposed in three steps: the white noise is first regularized introducing a momentum scale $\kappa$, then a field theory à la Janssen-De Dominicis-Martin-Siggia-Rose (JDMSR) \cite{janssen1976,dominicis1976,martin_siggia_rose,zinn} is employed to relate the calculation of observables to an effective $\kappa$-regularized action $\mathcal{S}_\kappa$, and finally a renormalization group (RG) analysis is led to define hypothetical fixed points in the RG flow procedure and to calculate the critical exponents of these fixed points. In our case, the white noise regularization can be done following Ref.~\cite{Nakayama2021EfimovTransition}, by taking the UV limit $\kappa \to 0$ of a spatially colored noise
\begin{equation}
    \langle \xi(\vec{x},t)\xi(\vec{x}',t') \rangle_\kappa = \delta(t - t')V_\kappa(\vec{x} - \vec{x}'),
\end{equation}
with $V_\kappa(\vec{r}=\vec{x} - \vec{x}')=(\kappa/\sqrt{\pi
})^3 e^{-(\kappa \vec{r})^2}$, such that $V_\kappa (\vec{r})  \xrightarrow[\kappa \to \infty]{} \delta(\vec{r})$. Interestingly enough, this noise regularization is extremely close to physical noise model build to reproduce flow temperature fluctuations in reactor physics \cite{Zylbersztejn2013OnPWRs}. Observables $ \langle \mathcal{O}[n] \rangle$ can then be expressed as functional integral over noise configurations through
\begin{equation}
\label{obs_mean}
     \langle \mathcal{O}(n) \rangle_\kappa = \int \mathcal{D}\xi(t) \mathcal{O}[n] P[\xi],
\end{equation}
where the normalization constant has been intentionally omitted, and which, still following the JDMSR approach, is shown to be equal to
\begin{equation}
\label{observable_final}
    \langle \mathcal{O}[n] \rangle_\kappa  =   \int \mathcal{D}n \mathcal{D}\Bar{n} \, \mathcal{O}[n] \exp^{-\mathcal{S}_\kappa [n, \Bar{n}]}.
\end{equation}
In this expression, the auxiliary field $\Bar{n}(\vec{x},t)$ has been introduced and the effective action $\mathcal{S}_\kappa$ is given by
\begin{equation}
    \mathcal{S}_\kappa[n, \bar{n}] = \int d\tau d\vec{x} \, \bar{n}(\vec{x},\tau) \left[ \partial_\tau - \Delta \right] n(\vec{x},\tau) - \frac{g}{4} \int d\tau d\vec{x} d\vec{x}' \bar{n}(\vec{x}, \tau) n(\vec{x},\tau) V_\kappa(\vec{x} - \vec{x}') \bar{n}(\vec{x}',\tau) n(\vec{x}', \tau)
\end{equation}
with a time re-scaling $\tau=2 D t$, and using the notation $g=\frac{\sigma^2}{D}$ for the bare coupling constant. Taking the UV limit and neglecting the power of the fields higher than quadratic (considered to be irrelevant under the renormalization group \cite{Nakayama2021EfimovTransition}), finally leads to 
\begin{equation}
    \mathcal{S}_\kappa[n, \bar{n}] \xrightarrow[\kappa \to \infty]{} \int d\tau d\vec{x} \, \Bigg[\bar{n}(\vec{x},\tau) \left[ \partial_\tau - \Delta \right] n(\vec{x},\tau) - \frac{g}{4} (n \bar{n})^2 \Bigg].
\end{equation} \\

\textbf{Roughening phase transition}
Introducing the dimensionless running coupling $\hat{g}=\frac{\sigma^2}{D}\kappa$, the renormalization of $\hat{g}$ is achieved by requiring this coupling to be $\kappa$ (i.e., cutoff) independent, which takes the form of fixed points in the dimension $d=3$ Calan-Symanzik equation
\begin{equation}
    \frac{\partial \hat{g}}{\partial \kappa} = \hat{g} - \frac{\hat{g}^2}{(4\pi)^{3/2}\Gamma(3/2)}.
\end{equation}
This equation possesses two fixed points. The fixed IR point $\hat{g}=0$ corresponds to $\kappa \to 0$, set in practice by the IR cutoff $\kappa^{-1}=L$. The fixed UV point $\hat{g}_c=(4\pi)^{3/2}\Gamma(3/2)$ corresponds to
$\kappa \to \infty$, obtained in practice for $\kappa^{-1}=a$. In coherence with References~\cite{Kardar1986DynamicInterfaces, Medina1989BurgersGrowth, Nattermann1992KineticRegime}, when considering numerical applications, we will set the minimum spatial scale $a$ close to unity \cite{Nattermann1992KineticRegime}, since $a=\zeta/\pi$, where $\zeta$ is the cross-over length \cite{Nattermann1992KineticRegime} below which no fluctuations are included \cite{Canet2011Non-perturbativeApplications}. As discussed in Subsection~\ref{sub:nrm}, in the power regime, the minimal spatial scale of fluctuations is considered to correspond to temperature perturbations with $\zeta \simeq 1$ cm \cite{Zylbersztejn2013OnPWRs}. Using these notation, the bare noise coupling $\sigma_{rt}$ corresponding to the roughening transition threshold can be expressed in terms of the bare diffusion coefficient of Eq.~\ref{eq_diff} following
\begin{equation}
\label{sigma_eq}
    \sigma_{rt}^2 = 4 \sqrt{\pi} D \, \zeta.
\end{equation}

\textbf{Renormalized parameters and critical exponents}
\label{rpce}
Below this neutron noise threshold, the behavior of the reactor should be given trustably by the mean field. Above this threshold, the RG flow pushes the coupling $\hat{g}$ and the coefficients of Eq.~\ref{eq_diff} toward renormalized values associated with the so-called strong noise regime. Therefore, the critical state within the reactor shifts to the KPZ universality class. The renormalized value $\hat{g}^*$ of the UV fixed point in dimension $d=3$ is shown to be close to $\hat{g}^* \simeq 5/2 \, \hat{g}_c$ \cite{Canet2011Non-perturbativeApplications}, which leads to a (spatial) scale dependent expression of the renormalized parameters $\sigma_\kappa$ and $D_\kappa$ given by 
\begin{equation}
\label{sigma_eq_eff}
    \sigma_\kappa^2 = 10 \pi^{3/2} \frac{D_\kappa}{\kappa}.
\end{equation}
When $\kappa^{-1}$ increases from the microscopic scale $a$ to the macroscopic scale $L$, the parameters acquire effective meaning: $\sigma_\kappa$ and $D_\kappa$ are, respectively, the observed/effective noise and diffusion coefficient at the spatial scale $\kappa^{-1}$. $\sigma_\kappa$ and $D_\kappa$ can be expressed in the form of power laws characterized by critical exponents
\begin{align}
\label{ce}
    & D_\kappa \sim  \kappa^{-(2-z)}, \\
\label{ce2}
    & \sigma_\kappa^2 \sim  \kappa^{-(3\chi+1)},
\end{align}
with $z+\chi=2$. Numerically, different estimates of critical exponents are found in the literature, such as $\chi=0.33$ \cite{RecentResultMultNoise}, $\chi=0.3$ \cite{Marinari2000CriticalSimulations} or $\chi=0.17$ \cite{Canet2011Non-perturbativeApplications} and in the following we will retain $\chi\simeq0.3$ thus fixing $z\simeq1.7$. From a reactor physics point of view, a scaling following $(\Delta x)^{z} \propto \Delta t$ with $z$ sensitively less than 2 tends to shift the random walk of neutrons from a diffusive behavior to a semi-ballistic one and might drastically change the spatial distribution of neutrons within the core, thus unambiguously signing its strong noise criticality regime. \\

\textbf{Rough neutron fields}
Also, the exponent $\chi$ describes the roughness $W_h$ of the interface in the KPZ equation, which scales as $W_h \sim L^\chi$, for sufficiently large $L$ and for large times. Although no equivalent exponent emerges from the study of the SHE equation with multiplicative noise, it is possible to estimate the 'roughness' of the neutron field under mild assumptions. Indeed, most nuclear reactors are considered decoupled in the sense that their typical size $L$ (few meters) is very large compared to the neutron's typical traveled path length before capture which is a few centimeters \footnote{The neutron's path length in a diffusive medium with diffusion with a (macroscopic) absorption cross section $\Sigma_a$ is given by $6\frac{D}{\Sigma_a}$ }, ensuring the validity of the $L^\chi$ scaling of the roughness of the KPZ interface. In the parametric sector where $\lambda<<\nu/h$ (which can be set by tuning $\lambda$ since the SHE is connected to a $\lambda$-parameterized family of KPZ equations, and assuming that h is bounded), the relation between the neutron field $n(\vec{x},t)$ and $h(\vec{x},t)$ becomes linear, thus ensuring a linearity (at first order) between the roughness of the neutron field $W_n$ and the roughness of the KPZ interface through $W_n=(\frac{\lambda}{2\nu})^2 W_h=(\frac{\lambda}{2\nu})^2 L^\chi$. The noise term $\eta$ is, however, enhanced by this re-parameterization (we have chosen a small non-linearity / high noise gauge), which acts as a prefactor but leaves the scaling law valid: at first order, the roughness of the neutron field should also scale as $L^\chi$. In the general case, independently of any particular approximation, and without calculating a precise scaling law, the neutron density field is by definition an exponential function of a potentially rough height field. This indicates that decoupled nuclear reactors develop in the power regime the equivalent of neutron clustering \cite{Dumonteil2021PatchyReactions} in the stochastic regime, and that the neutron field of the SHE also present a roughness if the roughening transition of the KPZ equation is attained.

\subsection{SHE equation with feedbacks}

\textbf{Multiplicative noise with a wall}
When the reactor power is increased, stabilizing feedbacks kick into action. The most prominent one is the Doppler broadening of the cross-sections: in thermal reactor, where neutrons are slowed down by elastic collisions in the moderator (water plus additional nuclei such as boron), the broadening of these cross-sections under temperature increase enlarges the heavy nuclei resonances and constitutes a trap for neutrons that are captured by the nuclei. Unable to participate in the chain reaction, the reactivity $\rho$ is sensitively diminished. Most models capture this effect using a term proportional to $-n^2$ \cite{BellG1970NuclearTheory,Akcasu1971MathematicalDynamics,Duderstadt1979TransportTheory}.
In~\cite{percolation_pre}, a stochastic noise model was built taking feedbacks into account. Paradoxically, this zero-power noise model of nuclear reactors adequately describes the maximal power attained by Monte Carlo simulations which, as of today, can simulate -in terms of neutrons- only a few Watts on modern parallel computing architectures \cite{Dumonteil2021PatchyReactions}. This model was therefore convenient for understanding simulation artifacts. Now, the same approach can be applied to the power noise model of a real reactor, given by Eq.~\ref{eq_diff} and supplemented by the $-\alpha n^2$ term. 
This model -of a stochastic heat equation with multiplicative noise plus powers of the field- has already been extensively studied in the literature, and, in the following, the main results demonstrated in Ref.~\cite{RecentResultMultNoise,munoz2003multiplicative} are recalled. Still through the Cole-Hopf transform, the stochastic equation with multiplicative noise
\begin{equation}
    \label{MN1}
    \frac{\partial}{\partial t} n(\vec{x},t) =  D \Delta n(\vec{x},t) + \sigma \xi(\vec{x},t) N(\vec{x},t) + \frac{\rho}{\Lambda } n(\vec{x},t)  -\alpha n(\vec{x},t)^2
\end{equation}
becomes the KPZ equation with a wall
\begin{equation}
\label{kpz_eq3}
    \frac{\partial}{\partial t} h(\vec{x},t) = D \Delta h + D \left( \vec{\nabla} h \right)^2 + \sigma \, \xi(\vec{x},t) + \frac{\rho}{\Lambda } -\alpha e^h.
\end{equation}  
This wall acts as a pinning structure constraining the evolution of the neutron density field, and the criticality of the reactor would correspond to a pinning/depinning phase transition. \\

\textbf{Common features with the KPZ equation}
A renormalization group analysis of the multiplicative noise equation (with a wall) and the simple KPZ equation (i.e. without a wall) however indicates that they share several common points \cite{Munoz2003MultiplicativeTutorial}: both have two fixed points separating a low noise Gaussian phase and a strong noise phase. Also, the transition point to the strong noise regime is still given by the same value of the dimensionless running coupling constant $\hat{g}_c$ calculated in Eq~\ref{sigma_eq} and, in the case of the strong noise phase, the critical exponent $z$ that governs the spatio-temporal scaling is also unchanged w.r.t. the KPZ equation, indicating an effective superdiffusive behavior of the neutron random walk. Consequently, $\chi$ should also be unaffected, as well as the $\kappa$-scaling of the effective parameters $\sigma_\kappa$ and $D_\kappa$. But supplementary critical exponents are associated with the pinning/depinning phase transition (via the presence of the wall), and the reader is refered to~\cite{RecentResultMultNoise,munoz2003multiplicative} for more information.


\section{\label{sec:noise} Reactor noise and instabilities}

We have seen that decoupled reactors might reach a new criticality regime dictated by the universality classes of the KPZ equation and the multiplicative noise SHE equation, for high noises given by equation Eq.~\ref{sigma_eq}. In this section, we will give more quantitative insight on the previous findings and we will correlate them with two particular events that exhibited strong noise patterns and power instabilities.

\subsection{Phenomenology}

More than 430 nuclear reactors are currently in operation worldwide, the vast majority of them being thermal water-moderated reactors: neutrons are slowed down through successive scattering on protons and induce fissions on heavy nuclei once thermalized. Among these reactors, the two main technologies are pressurized water reactors (PWR) and boiling water reactors (BWR). While both technologies benefit from a long and extensive experience feedback -dating back to the 1970s- neutron flux fluctuations measured by in-core and ex-core instrumentation have attracted much attention for two opposite reasons. On the one hand, neutron noise is a powerful diagnostic tool for probing a reactor during operation and even to characterize and sometimes locate perturbations in the core. On the other hand, large noise baselines in PWR reactors and instability events with sometimes sudden increases in noise in BWRs have themselves been a recurrent cause for concern. Originally termed reactor instabilities, the phenomenology of these largely unexplained sudden neutron noise patterns has soon caught the attention of reactor physicists (see, for instance, the OECD international benchmark on the Swedish BWR Ringhals 1 core \cite{Lefvert1995RinghalsBenchmark}). Since then, several such events have occurred in BWRs, noticeably in Sweden and Mexico \cite{Olvera-Guerrero2017Non-LinearEntropy}. The attention devoted to neutron noise has recently been renewed, taking the form of international collaborations \cite{Verma2020AssessmentReactor}, as recent measurements in various PWR reactors have identified drifts and contained increases of their 'neutron noise' during operations. Furthermore, many of these reactors exhibit a high baseline of noise during normal operation, such as in pre-Konvoi and Konvoi PWRs in Germany \cite{Seidl2015ReviewPWRs}, Switzerland \cite{Girardin2018DetailedGoesgen} or Spain \cite{2012AlmarazCO-12/043.,Zylbersztejn2013OnPWRs}, and even more recently in the first operating EPRs \cite{Jiang2022InvestigationEPR}. Although normal noise levels depend on reactors (they are close to $0.2\, \%$ \footnote{in terms of standard deviation of the core power} for Westingouse PWRs, $2\, \%$ for Konvoi and $5\, \%$ for preKonvoi \cite{Bermejo2017OnS3K}), both progressive increase exceeding $10\, \%$ \cite{Viebach2021InvestigationsReactors} and sudden unstable events developing on $\sim1$ minute time scale have been recorded, particularly on BWRs \cite{Olvera-Guerrero2017Non-LinearEntropy, Gavilan-Moreno2016UsingReactors}. In these last cases, fluctuations would swiftly exceed these values, leading to immediate reactor scramming \cite{Torres2019NeutronCore}. Depending on local regulations and industrial procedures, some reactors now have a reduced nominal power to avoid the development of such instabilities, with a threshold usually set between 8 and 10$\%$ \cite{Bermejo2017OnS3K}. In addition, as said in the previous section, many reactor automatic protection systems are based on the calculation of the time derivative of the core power to eventually prevent unwanted power excursion. This strategy can be affected by high neutron noise, thus translating a commercial matter into a nuclear safety concern.

The state of the art in the understanding of the neutron noise relies on the analysis of the characteristic frequencies at which fluctuations occur. The typical frequency spectrum is different on BWRs and PWRs: while peaked at frequencies below 1 Hz (often close to 1/2 Hz), the spectrum is rather flat for BWRs, which develop stronger and more frequent noise events-, and the white noise hypothesis is hence more realistic. In contrast, the spectrum of PWRs is closer to a decaying exponential (still in the frequency domain) \cite{Rouchon2016AnalyseRapides}. For both reactors, fluctuations can be either out-of-phase (or regional) whenever they have opposite sign in different regions of the core (e.g. upper/lower part of the core) or in-phase (or global) when they are synchronized throughout the core. This militates in favor of a C- or S-shaped vibration of the assemblies \cite{Demaziere2022UnderstandingTheory}, connecting the neutron noise to mechanical vibrations of the core components between 1 Hz and 50 Hz (of fuel rods and assemblies and thermal shield below 25 Hz, and of the reactor vessel and internals below 50 Hz). Below 1Hz, they are believed to be connected to temperature fluctuations in the primary loop. The two mechanisms have different control pathways on the neutron population: in the case of coolant temperature fluctuations, both density and Doppler effect affect the probability to slow down neutrons and hence their capacity to reproduce through fissions, while in the case of mechanical vibrations, it is the inter-assembly coolant thickness that is modified.   

In the next two subsections, a neutron noise model relying on the strong noise hypothesis will be built, and correlated on two sets of experimental data -one corresponding to a PWR and the other to a BWR-, so as to grasp at a minimal cost the wide variety of neutron noise events.

\subsection{Almaraz Trillo PWR noise data}
\begin{figure}[!ht]
\centering
\includegraphics[scale=0.55]{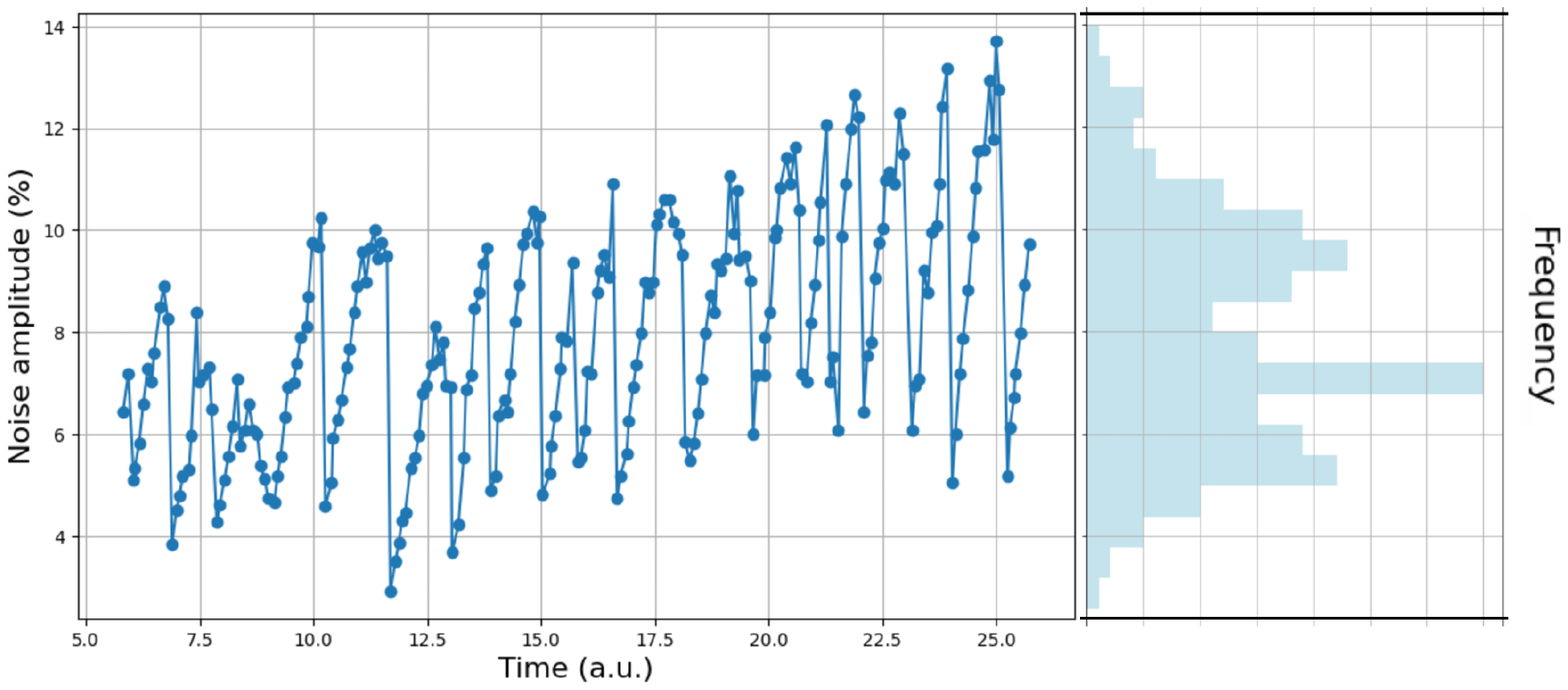}
\caption{Almaraz-Trillo PWR data from \cite{Kraus2011ConsequencesPwrs, Viebach2021InvestigationsReactors}. Left plot: RMS of the neutron noise amplitude (in $\%$) as a function of time (in arbitrary units). Local minima correspond to BOC measurements and local maxima to EOC measurements. Right plot: noise amplitude histogram of all measurements. The estimated average noise is $8\%$.}
\label{sig_vs_t}
\end{figure}

Being related to commercial issues, data released from power plants in general and concerning neutron noise in particular are relatively scarce. Among published data available, experimental measurements associated with increased noise levels from the 1000 MW Almaraz Trillo PWR reported in 2011 \cite{Kraus2011ConsequencesPwrs, Viebach2021InvestigationsReactors} present an interesting scheme, as can be observed in Figure~\ref{sig_vs_t}, giving the time evolution of the root mean square of the neutron noise measured by ex-core detectors (left plot), as proposed by these references, and a projection of the noise amplitudes in a histogram aiming at presenting both the average noise and its typical deviations (right plot). In fact, throughout the cycle, this noise is progressively amplified by a factor of 2 between cycles 8 and 25, with intracycle variations ranging at their extremum from 4 to 14$\%$. In this section, this data sample will be used for comparison with the roughening phase transition model derived above, looking first at the overall amplitude of the noise and then at the variations of the noise.

At cycle 20, no more general trend seems to affect cycle-to-cycle fluctuations, and the noise response of the reactor seems stationary. A histogram that reports the noise amplitude from that cycle is shown in Figure~\ref{sig_vs_t}. Despite large fluctuations in the noise itself, which can be attributed to intracycle effects \footnote{So as to compensate the reactivity loss due to fuel burnup effects, boron is used at beginning of cycle in an amount adjusted in such a way that the reactivity through the cycle stays constant. However, the average absorption cross sections of the moderator decrease linearly with time during the cycle, in the first order, leading to the observed linear increase in noise induced by the cross sections}, the average noise value extracted from this graph is estimated to be close to 10$\%$. \\

\textbf{Noise amplitude from the roughening transition}
Making the assumption that reactor instabilities are related to the roughening transition, the noise threshold at which the transition occurs can also be numerically evaluated, in a context where spatial patterns of noise, while presenting involved structures, as analyzed, for example, in the linear noise theory ~\cite{Demaziere2022UnderstandingTheory}- are neglected. The noise given by Eq.~\ref{sigma_eq} is associated with Eq.~\ref{eq_diff} which, once solved, gives access to the neutron density field $n(\vec{x},t)$. Closely following the approach detailed in~\cite{2010HandbookEngineering}, and since we are only interested in temporal fluctuation of a spatially averaged noise, Eq.~\ref{eq_diff} can be rewritten factorizing the purely time-dependent amplitude $P(t)$ out of a remaining spatial part of the neutron density that is considered constant $n(x,t)\approx n_0 $. Eq.~\ref{eq_diff} hence takes the simplified form
\begin{equation}
\label{eqpt}
    \frac{\partial}{\partial t} P(t) = \frac{\rho}{\Lambda} P(t) + \sigma \xi(\vec{x},t) P(t).
\end{equation}  
In addition, the spatiotemporal noise term $\eta(\vec{x},t)=\sigma \xi(\vec{x},t)$ and its intensity $\sigma$ were implicitly related through the definition
\begin{equation}
    \langle \eta(\vec{x},t) \eta(\vec{x}',t') \rangle  = \sigma^2 \delta^{3}(\vec{x}-\vec{x}') \delta(t-t'),
\end{equation}
which spatial dependence can readily be integrated over the whole reactor of volume $V=\int dx^3$ still assuming that we are only interested in the temporal component of the noise. Therefore, the resulting temporal noise takes the form
\begin{equation}
    \langle \eta(t) \eta(t') \rangle  \int dx^3 = \sigma^2 \delta(t-t') \int dx^3 \delta^{3}(\vec{x}),
\end{equation}
\begin{equation}
    \langle \eta(t) \eta(t') \rangle = \frac{\sigma^2}{V} \delta(t-t'),
\end{equation}
hence recasting equation Eq.~\ref{eqpt} in a classical white noise equation
\begin{equation}
\label{eqptf}
    \frac{\partial}{\partial t} P(t) = \frac{\rho}{\Lambda} P(t) + \frac{\delta \rho }{\Lambda} P(t)
\end{equation}  
(where we have used the notation $\delta \rho / \Lambda = \frac{\sigma}{\sqrt{V}} \xi(t)$ for the volume-rescaled noise coupling). To derive an equation for the noise from this equation as suggested by Ref.~\cite{2010HandbookEngineering}, $P(t)$ is also split in a (time) fluctuating component $\delta P(t)$ plus a constant term $P_0$. Second order terms are neglected, and the system is considered to be close to criticality so that one obtains the equation for the fluctuations
\begin{equation}
\label{eqptf2}
    \frac{\partial}{\partial t} \delta P(t) =  \frac{\delta \rho }{\Lambda}  P_0 = \frac{\sigma}{\sqrt{V}} \xi(t) P_0,
\end{equation}  
or equivalently in the frequency domain, after a Fourier transform
\begin{equation}
\label{eqptf3}
    \Big| \frac{\delta P(w)}{P_0} \Big| = \frac{\sigma^2}{V w}.
\end{equation}  
This equation connects the fluctuations in the amplitude factor of the neutron density, i.e., the neutron noise $\delta P/P_0$, to the reactivity noise $\delta \rho$, which was formulated as a function of the noise coupling term $\sigma$. Following Eq.~\ref{sigma_eq}, the neutron noise at which the roughening transition occurs, noted $\Big| \delta P/P_0 \Big|_{rt}$, is given by
\begin{equation}
\label{eqptf4}
    \Big| \frac{\delta P}{P_0} \Big|_{rt} = \frac{\sigma_{rt}^2}{V w}= 4 \sqrt{\pi} \frac{D \zeta}{V w}.
\end{equation}  
A numerical estimation of this term can be worked out by setting the observation frequency $w$ to $1$Hz - close to its maximum- and using a cylindrical reactor volume of typical radius $r=150 ~cm$ and $H=400 ~cm$ to calculate $V$. The bare diffusion coefficient $D$ is set to $10^5 ~cm^{2}.s^{-1}$ as suggested by the thermal energy PWR data of~\cite{Zylbersztejn2013OnPWRs}. Also in accordance with this reference, and as stated in the previous section, $\zeta \simeq 1$ cm is the minimal spatial scale below which no power fluctuations occur and can be associated with a temperature-induced change of the fuel absorption cross section. For these parameters, the RMS of the noise amplitude is $\sqrt{\Big| \delta P/P_0 \Big|_{rt}} \simeq 15\%$. 
As can be observed on the right plot of Figure~\ref{sig_vs_t}, the experimental data present a bimodal distribution, the noise amplitude progressively drifting within the cycle from a local minimum value to a local maximum value. The average value (over all available cycles) is, as discussed in the previous paragraph, close to $10\%$. Given the level of detail of the roughening transition model, its value can be considered to be very close to these experimental data. Indeed, any increase/decrease in the minimal correlation length $\zeta$ linearly affects the noise threshold estimation. But this length differs sensitively from few mm to few cm depending on scarce experimental measurements and widely varying theoretical modeling \cite{2010HandbookEngineering,Zylbersztejn2013OnPWRs}. Beyond this agreement, it should be noted that Eq.~\ref{eqptf4} sets the dependence of the noise at which the transition occurs on other physical parameters: the bigger the reactor volume $V$ -or the smaller the diffusion coefficient $D$-, and the smaller the noise (i.e. the vibration level) at which the reactor meets the rough regime. This prediction could also lend itself to experimental confirmation.

In this subsection, we considered a -close to- stationary reactor with a long-term high noise level and showed that its noise amplitude was compatible with the noise associated with the roughening transition. To better assess a potential switch from a Gaussian behavior to a rough behavior in a nuclear reactor, the ideal situation would be to consider a hypothetical event where the reactor switches, on small time scales, from standard operating conditions (assumed to belong to the Gaussian fixed point) to a degraded state of the reactor with a high neutron noise level (that could be correlated to the strong noise regime) and to try to associate the shift in the critical parameters to the universal exponents of the phase transition. Such situations of abrupt neutron noise increase are more frequently met with BWRs and therefore, in the next subsection, the Laguna Verde BWR incident is discussed.

\subsection{The Laguna Verde emergency incident}

\textbf{BWRs}
Noisy baselines and instability events occur more frequently in BWRs than in PWRs and also seem to develop on smaller time scales or equivalently with heavier tails in the frequency spectra. By comparison to PWR, BWR have a slightly harder neutron spectrum: a proper modeling of their noise would fairly require the so-called two-group approximation, where two coupled noisy diffusion equations should be used for each energy group. But their particularity is believed to be associated with the two-phase coolant flow: liquid water is injected at the bottom of the reactor and subject to a liquid-vapor phase transition during its travel through the core. The resulting scale-free structure of bubble sizes during the boiling crisis has recently been experimentally put into evidence and shown to belong to the percolation universality class \cite{Zhang2019PercolativeCrisis}. The minimal size of the bubbles is considered to be $\approx 0.01$ cm \cite{Alvarez-Ramirez2005DetrendedReactor,Uga1972DeterminationBWR} and defines the minimal correlation length $\zeta$, which, recalling Eq.~\ref{eqptf4}, lowers the threshold in the noise amplitude associated with the roughening transition from an order of magnitude compared to PWR, thus eventually occurring at $\sqrt{\Big| \delta P/P_0 \Big|_{rt}}\simeq 1.5\%$. This is relatively consistent with empirical observations that relate more frequent and abrupt noisy events in BWR than in PWR. \\

\begin{figure}[!ht]
\centering
\includegraphics[scale=0.7]{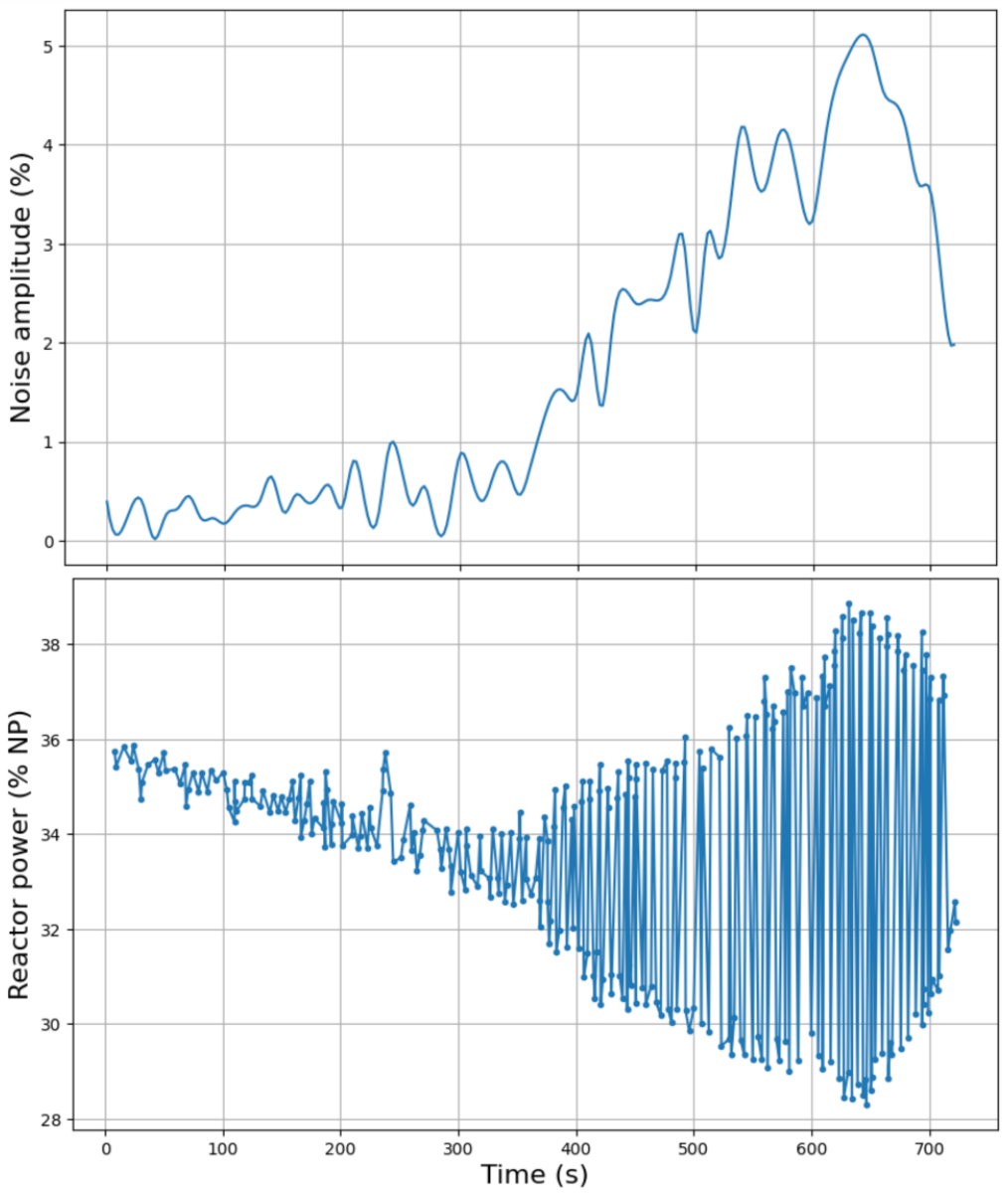}
\caption{ Bottom plot: experimental data of the Laguna Verde BWR 1995 instability event extracted from \cite{Blazquez2003TheLearnt,Alvarez-Ramirez2005DetrendedReactor, Olvera-Guerrero2017Non-LinearEntropy}, presenting the average power level versus time, as recorded by APRM detectors. Top plot: linear interpolation of the neutron noise amplitude during the event.}
\label{fig4:LagunaVerde1}
\end{figure}

\textbf{The event}
To provide more quantitative insight into such phenomena and to illustrate such an event occurring on short time scales, the 1995 Laguna Verde emergency incident is now considered \cite{Blazquez2003TheLearnt,Alvarez-Ramirez2005DetrendedReactor,Blazquez2013SearchingTransform,Olvera-Guerrero2017Non-LinearEntropy}. It happened in January 1995, during the start-up of the reactor, and lasted 723 s during which power oscillations grew apparently unbounded, even though the power was slightly decreased in an attempt to control their amplitude (see bottom plot of Figure~\ref{fig4:LagunaVerde1}, where an extraction of visible data points from the previous references has been performed). As the maximum peak-to-peak value of the oscillations reached $10\%$, a scramming procedure was triggered. The top plot of Figure~\ref{fig4:LagunaVerde1} presents the growth of the noise during the event, using an interpolation of the raw data after subtracting the average decrease trend in power. As can be observed, the initial noise amplitude was below $1\%$, indicating a stable behavior of the reactor, but initiated an apparent exponential growth with a period close to 100 s.\\

\begin{figure}[!ht]
\centering
\includegraphics[scale=0.5]{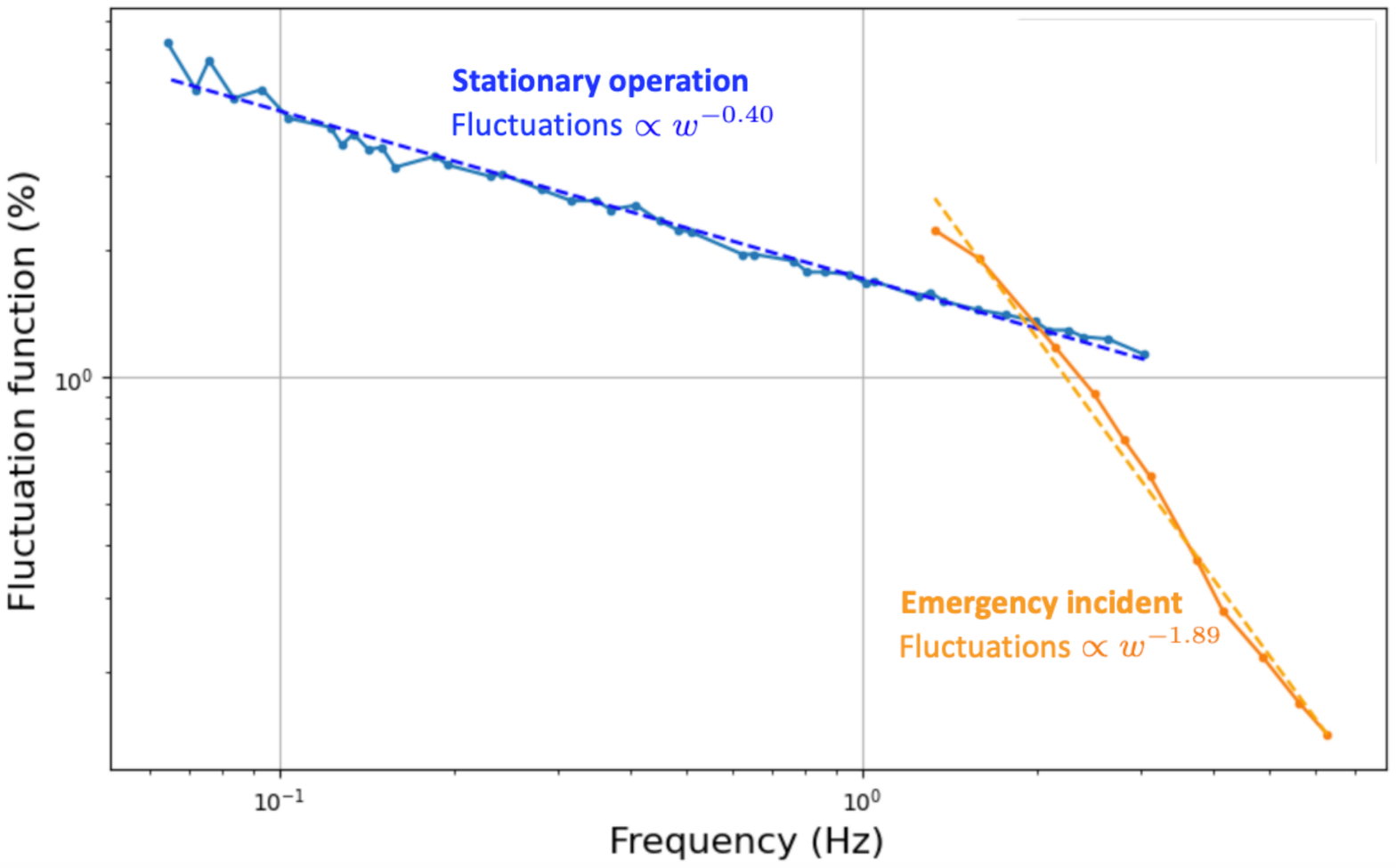}
\caption{Experimental data of the Laguna Verde BWR 1995 instability event \cite{Blazquez2003TheLearnt,Alvarez-Ramirez2005DetrendedReactor, Olvera-Guerrero2017Non-LinearEntropy}. Fluctuation function as proposed by Reference~\cite{Alvarez-Ramirez2005DetrendedReactor} presented versus frequency $w$, during a stationary operation of the Laguna Verde reactor (data: blue solid line, fit: blue dashed line) and during the emergency incident (data: orange solid line, fit: orange dashed line).}
\label{fig5:LagunaVerde2}
\end{figure}

\textbf{Shift in power spectra due to roughening transition}
The roughening transition to a strong noise regime could once again provide information on this particular phenomenology. Indeed, the scale-free behavior of the boiling \cite{Zhang2019PercolativeCrisis} discussed above mirrors a scale-free correlation length $\kappa^{-1}$ and could enforce the scale invariance hypothetized by the renormalization procedure. As discussed before, the roughening transition threshold is lower due to the minimal size of the bubbles, explaining that BWRs might be proner to enter a rough noise regime. In the case of Laguna Verde, the tripping of the reactor has occurred during a change of equilibrium where the power was rather constant on average (it was initially lowered during the first stage of the incident), but where the noise amplitude was promptly increasing. Unlike the PWR case in the previous subsection, where a strong noise equilibrium was hypothetized, the emergency procedure most probably prevented such a regime (if any) in Laguna Verde. Since noise amplitudes characteristic from this regime cannot consequently be estimated, interpreting this change of the behavior of the reactor requires to use other tools, such as a frequency spectrum analysis. Eq.~\ref{eqptf3} can be rewritten 
\begin{equation}
\label{eqbwr}
    \Big| \frac{\delta P(w)}{P_0} \Big| = \delta \rho(w) G_0(w),
\end{equation}  
where the reactivity perturbation $\delta \rho$ is defined as before, and where the $G_0(w)=\frac{1}{w \Lambda}$ is the so-called zero-power reactor transfer function \footnote{which is modified as $G_0(w)=\frac{1}{w*(1+\frac{\beta/\Lambda}{iw+\lambda})}$ when delayed neutron are taken into account, with $\beta$ the delayed neutron fraction and $\lambda$ the precursor decay rate}. For more realistic transfer functions that take into account complex space-dependent thermal-hydraulic and neutronic phenomena, the differential equation Eq.~\ref{eqptf} is sensitively modified, and the effective transfer function $G(w)$ could present substantial deviations from $G_0$. In any case, auto-power spectrum density (APSD) signals are often characterized by power laws for the $w$ spectrum, which allow one to understand rapid kinetic changes in the reactor.
However, trying to appreciate the hypothetical change in APSD power laws associated to the roughening transition can be based on the fact that the change from normal behavior to a strong noise regime is only dictated by the shift in the tails of $\delta \rho(w)$. For a white noise signal,  $\delta \rho$ is constant. This corresponds to the temporal white noise $\eta(\vec{x},t)$ associated with the bare noise coupling $\sigma$. In the strong noise regime, $\sigma_\kappa$ acquires an effective meaning and depends on the observation scale $\kappa^{-1}$: the strength of the effective noise coupling is crucially dependent on this scale. The scale-dependent change in the effective noise amplitude is given as a power law associated to a critical exponent and, using Eq.~\ref{ce2}, takes the form
\begin{equation}
\label{eqkappa}
    \sigma^2(\kappa)\propto \kappa^{-\eta_D^*},
\end{equation}  
with $\eta_D^*=3\chi+1$ and $\chi\simeq 0.3$ (see Subsection~\ref{rpce}). 
Now, since perturbations in the case of BWR are believed to be mainly correlated with void effects arising from the two-phase flow, a simplified model can be postulated that inversely relates the frequency scale $w$ at which the reactor is observed to the spatial scale $\kappa^{-1}$ at which it is observed. In fact, the coolant transports perturbations at a constant velocity $v \approx 100$ cm.s$^{-1}$. At a given position, the duration associated with a perturbation is hence given by $w^{-1}=2\pi\kappa^{-1}/v$, from which the proportionality between $w$ and $\kappa$ is hypothesized. In this model, a small perturbation of $\kappa^{-1} \simeq 1$ cm corresponds to 10 Hz, whereas a quarter-core perturbation of $\kappa^{-1} \simeq 100$ cm corresponds to 1 Hz: these perturbations will define the dynamic frequency range for the observations. Finally, it is possible to appreciate the change in the spectrum of the reactor transfert function $G(w)$ when the core shifts from normal operating conditions to a strong noise regime, using
\begin{equation}
\label{eqbwr2}
    \Big| \frac{\delta P(w)}{P_0} \Big| \propto \frac{G(w)}{w^{3\chi+1}}.
\end{equation}  
Therefore, the RMS of the noise amplitude should approximately scale as $w^{-1/2}$ under normal conditions (with $G(w)=G_0(w)\propto 1/w)$), while the scaling associated with a strong noise regime should be given by $w^{-(1/2+(3\chi+1)/2)} \approx w^{-3/2}$.
To appreciate whether such a change had occurred during the Laguna Verde incident, data were reanalyzed according to the methodology of Reference~\cite{Alvarez-Ramirez2005DetrendedReactor}. This paper compares a standard sequence of recorded data of Laguna Verde at nominal power (NP), to the data recorded during the incident. Since the reactor was operated at $30\%$ NP at the outbreak of the incident and, furthermore, during a power decrease, Alvarez-Ramirez et al. proposed a detrended noise analysis of the sequence. The results of the analysis, consistent with Reference~\cite{Alvarez-Ramirez2005DetrendedReactor}, appear in Figure~\ref{fig5:LagunaVerde2}, where the fluctuation values of the experimental data are extracted from Reference~\cite{Alvarez-Ramirez2005DetrendedReactor} but presented as a function of the frequency $w$ and adjusted only by focusing on large values $w$ (over a 700 s signal). The blue points represent the fluctuation values associated with the Laguna Verde reactor operating under normal conditions (the $G(w)$ function), while the orange data points are experimental data recorded during the incident. The noise amplitude estimated by the detrended fluctuation function clearly follows a different $w$ scaling. The results of the adjustment using power laws of both signals are indicated in the plot. The normal sequence, labeled "stationary operating conditions", indeed exhibits a frequency power law of 0.4, close to the $1/\sqrt{w}$ associated with a white noise signal. This anticorrelation most probably signs the presence of feedbacks that keep the signal stationary \cite{Alvarez-Ramirez2005DetrendedReactor}. Concerning the incidental signal, labeled "emergency incident", the fit of the slope on a log-log scale is also coherent with \cite{Alvarez-Ramirez2005DetrendedReactor}, and equal to $1.89$ with a $2\%$ confidence interval, when observed on the dynamic range of $[1-10]$ Hz. This value is sensitively higher than the value calculated for the strong noise regime but still coherent given the level of simplifications of the model. It should also be noted that $1/w^2$ noise spectra were also characterized in this range for the APSD of high-noise events observed in PWRs \cite{Seidl2015ReviewPWRs}, which could indicate that the insights provided by the strong noise-induced shift in the frequency power law calculated for BWRs could also contribute to understanding open issues of PWR noise measurements. 

\section{Conclusion}

Small and limited power fluctuations are key to the safety of nuclear reactors. On the one hand, automatic protection systems rely on them to assess the reactor power time derivative and take appropriate actions. On the other hand, neutron noise measurements are sometimes used to assess the condition of operating reactor cores. However, for reasons that are only partially understood using linear noise theory (on which both theoretical and numerical modeling of neutron power noise are based), the fluctuations themselves can sometimes increase to unusual levels, as in some PWRs, or grow unboundedly, as has been observed in some BWRs, and hence pose direct safety concerns. 

In this paper, exploiting the formal proximity between such reactor fluctuations models and the stochastic heat equation, a field-theoretic approach to the temporal and spatial behavior of reactors has been proposed. Through a Cole-Hopf transformation, the one-group diffusion equation with stochastic multiplicative noise is indeed mapped onto the KPZ equation, plus a wall if feedback mechanisms are taken into account. This observation paved the way for the study of neutron power noise using renormalization techniques, which predicted that, under certain conditions, the reactor could switch to a new criticality state, profoundly different from standard criticality, and that the neutron fields could become rough, in direct analogy to the dynamics of surface growth, even when feedback mechanisms are taken into account.

The phenomenology of such rough neutron fields was correlated with two specific events. First, the noise amplitude measured during a noise increase of the Almaraz-Trillo PWR was compared to the noise threshold of the roughening transition. The proximity of both values thus confirmed the relevance of interpretations of PWR high noise sequences with the help of the strong-noise universality class. Second, in BWR, the small minimal correlation length related to the two-phase flow was shown to lower this threshold in the noise amplitude down to values currently met during operations, hence eventually explaining why these reactors are proner to develop high noise events. Data from the Laguna Verde BWR accident in 1995 were used to characterize the spectral frequency shift that separates the standard operation of the reactor from its incidental functioning. This shift was also evaluated from the roughening transition model, showing a good agreement with the experimental value.

From the perspective of reactor physics, this theory, if further supported by data, could help characterize and understand instabilities, especially those occurring in BWRs, and hopefully design diagnostic tools to prevent their appearance. For example, the use of largest Lyapunov exponents has been proposed \cite{Gavilan-Moreno2016UsingReactors} in the context of real-time analysis of BWR plant data, which typically implies fractional exponents similar to those derived using the roughening transition model. This seems to be all the more feasible as the Lyapunov exponents of the SHE with multiplicative noise have recently been derived, albeit in lower dimensions \cite{Ghosal2023LyapunovData}. For PWRs, long-term trends (over many cycles) in noise amplitude have been measured in various reactors. Such phenomena may also benefit from interpretation in light of the roughening phase transition. Indeed, if, as has been calculated, temperature fluctuations and core component vibrations put the reactor close to the phase transition limit, any small-scale persistent perturbation would explain such long-lasting trends.

Finally, on a different note, from the perspective of statistical physics, BWR noise events could represent a first measurement of critical exponents of the KPZ universality class in dimension $d=3$, thus confirming the adage that every cloud has a silver lining.


\section*{Acknowledgments}

The author is grateful to Benjamin Dechenaux and Léonie Canet for very helpful comments and illuminating discussions. 

\bibliography{biblio,mendeley}

\end{document}